\documentclass[runningheads]{llncs}

\usepackage{graphicx}
\graphicspath{{./figures/}}
\usepackage{booktabs} 
\usepackage{tabu}
\usepackage{paralist}
\usepackage[skip=0pt]{caption}
\usepackage{acronym}
\usepackage{enumitem}
\usepackage{amsfonts}
\usepackage{hyperref}
\usepackage{amsmath}
\usepackage{amsfonts}

\usepackage[square,comma,numbers,sort&compress,sectionbib]{natbib}
\usepackage{soul}

\usepackage{subfig}

\usepackage{tikz, tikzscale} 
\usetikzlibrary{shapes.geometric,arrows,calc,positioning,decorations.pathreplacing,bending}

\usepackage{algpseudocode}
\usepackage{algorithm}
\algrenewcommand{\algorithmiccomment}[1]{// #1}
\algnewcommand\algorithmicinput{\textbf{Input:}}
\algnewcommand\Input{\item[\algorithmicinput]}
\algnewcommand\algorithmicoutput{\textbf{Output:}}
\algnewcommand\Output{\item[\algorithmicoutput]}

\acrodef{LTR}{learning to rank}
\acrodef{OLTR}{online learning to rank}
\acrodef{DNN}{deep neural network}
\acrodef{DCG}{discounted cumulate gain}
\acrodef{MAB}{multi-armed bandits}
\acrodef{MGD}{multileave gradient descent}
\acrodef{DBGD}{Dueling Bandit Gradient Descent}

\newcommand{\OurMethod}{RegLearn}
\newcommand{\OurMethodPG}{PGLearn}
\newcommand{\skyline}{OracleLearn}

\title{Online Learning to Rank with List-level Feedback for Image Filtering }
\author{Chang Li\inst{1} 
	\and Artem Grotov\inst{1}\thanks{Now at GoDataDriven, the Nehterlands.}
	\and Ilya Markov\inst{1} 
	\and \\ Maarten de Rijke\inst{1}
}
\authorrunning{C. Li et al.}
\institute{University of Amsterdam, Amsterdam, the Netherlands\\
	\email{\{c.li, i.markov, derijke\}@uva.nl}\\
		\email{agrotov@gmail.com}
}

\allowdisplaybreaks
\begin{document}

\maketitle

\begin{abstract}
Online learning to rank (\acsu{OLTR}) via implicit feedback has been extensively studied  for document retrieval in cases  
where the feedback is available at the level of individual items.
To learn from item-level feedback, the current algorithms require certain assumptions about user behavior. 
In this paper, we study a more general setup: \ac{OLTR} with list-level feedback, where the feedback is provided only at the level of an entire ranked list. 
We propose two methods that allow online learning to rank in this setup. 
The first method, \OurMethodPG, uses a ranking model to generate policies and optimizes it online using policy gradients.
The second method, \OurMethod, learns to combine individual document relevance scores by directly predicting the observed list-level feedback through regression.
We evaluate the proposed methods on the image filtering task, in which \acp{DNN} are used to rank images in response to a set of standing queries.
We show that \OurMethodPG{} does not perform well in \ac{OLTR} with list-level feedback.
\OurMethod, instead, shows good performance in both online and offline metrics.
\end{abstract}


\section{Introduction}

Image search concerns a large portion of modern  search engine systems~\cite{xie-why-2018,xie-investigating-2017}. 
In this paper, we focus on the image filtering task, a special case of image search,   which aims at identifying relevant images given a standing query~\cite{filtering1,filtering3}. 
We are particularly interested in scenarios that are characterized by a fixed set of information needs and a set of images that needs to be ranked against each of those needs, e.g., video surveillance with a fixed visual vocabulary~\citep{boom-research-2014}, (visual) reputation monitoring~\citep{amigo-overview-2014}, or visual information discovery services such as Pinterest,\footnote{\url{http://www.pinterest.com}} etc.

How can we use state-of-the-art \acf*{OLTR} methods to learn to improve performance on the image filtering task?
The image filtering task poses a number of challenges for modern \ac{OLTR} methods. 
First, previous \ac{OLTR} methods are mostly based on a typical text search setup, where features are textual similarities between query  and document, e.g., BM25 and TF-IDF~\cite{ltr}.  
How to learn and rank images by pixel features online is still an open question.
Second, in many specific instances of the image filtering task, the feedback from which one has to learn is not available at the level of individual items but only at the level of the entire result list~\citep{xie-2018-image}.
Think, in particular, of scenarios such as user satisfaction in mobile search, or with intelligent assistants, where feedback comes from a potentially diverse set of user interactions, e.g., user gestures, search dialogues~\cite{Williams:top5,kiseleva2016predicting}, and where large volumes of feedback can only be acquired at the level of an entire ranked list of items.

More specifically, then, the problem that we address in this paper is how can we develop \ac{OLTR} methods for image filtering that learn from list-level feedback?
In \ac{OLTR}, we aim to constantly improve the underlying ranking model based on an incoming stream of user feedback such as clicks.
A range of \acf{OLTR} techniques have been proposed in the literature~\cite{dbgd,rankbandit,balancing,mdgd,batchrank,kveton2018bubblerank}.
The methods proposed so far only operate with item-level feedback, such as knowing whether a particular item is clicked or not, which allows us to assess the quality of an individual item.
Since user feedback is biased~\cite{yue2010beyond,joachims2017accurately}, to get rid of the bias, \ac{OLTR} algorithms with item-level feedback either require additional assumptions, e.g., the cascading assumption~\cite{kveton2015cascading}, or a certain level of randomization~\cite{batchrank,kveton2018bubblerank}. 
As a result,  algorithms of the first group are limited by the assumptions they require, and  algorithms of the second group potentially hurt the user experience during the early stages of learning~\cite{kveton2018bubblerank}. 

Unlike item-level feedback, list-level feedback measures the quality of ranked lists directly,  such that a ranked list of higher quality receives more clicks on average than a ranked list of lower quality. 
Learning from  list-level feedback allows the algorithm to avoid debiasing procedures used by item-level based algorithms. 
And, clearly, item-level feedback can be turned into list-level feedback (but not vice versa). 
Meanwhile, from the point of \ac{LTR}, list-wise methods generally performs better than point-wise methods~\cite{li-incremental-2018,ltr}.

The challenge of list-level \ac{OLTR} is that a ranked list consists of multiple items (e.g., $5$ images), but the obtained feedback is a single value (e.g., abandonment). 
To learn a ranking model in this scenario, we need to compute the contribution of individual items to the list-level feedback.
We propose two methods to do so.
The first method, \OurMethodPG, takes a policy gradient point of view and considers each ranked list as an action and optimizes the policy that chooses the best action given a user's query based on the observed reward (i.e., list-level feedback).
The second method, \OurMethod, uses regression to directly learn to combine individual item relevance scores to predict the observed list-level feedback.
Then, \OurMethod\ employs back-propagation to update the underlying ranking model that produces the relevance scores.
In order to learn from pixel features, we choose a \acf{DNN} as the underlying ranking model for both \OurMethod{} and \OurMethodPG{}. 

We simulate the image filtering task on the MSCOCO dataset~\cite{mscoco} and consider nDCG@$k$ (the ideal setup) and CTR@$k$ (the noisy setup) as list-level feedback. 
Our experimental results show that \OurMethodPG{} performs poorly when the list size is larger than $2$ while \OurMethod{} is able to train a good underlying ranking model with larger lists.

In summary, the main contributions of this paper are the following:
\begin{enumerate}[nosep]
	\item We propose two methods, \OurMethodPG\ and \OurMethod, which allow online learning of ranking models in scenarios where only list-level feedback is available.
	\item We evaluate the proposed methods on the image filtering task and show that \OurMethod\ has superior performance compared to \OurMethodPG, especially when the list size increases.
\end{enumerate}


\section{Related work}
\label{section:related-work}

\paragraph{Online  learning to rank.}
\label{sec:oltr}
Most previous work on \ac{OLTR} either formulates the \ac{OLTR} problem as a \ac{MAB} problem~\cite{rankbandit,batchrank,kveton2015cascading,kveton2018bubblerank} or as a  \ac{DBGD} problem~\cite{dbgd,balancing,mdgd,wang2018efficient}. 
MAB-type algorithms rank items by an item-wise estimator that estimates the probability of an item being clicked. 
This type of algorithms only use item-level feedback and are not generalized across different queries. 

\ac{DBGD} has been proposed by~\citet{dbgd} and learns a ranking function by gradient descent via interleaved comparisons. 
\citet{mdgd} extend \ac{DBGD} to learn from multileaved comparisons and propose a more effective algorithm: Multileave Gradient Descent. 
Since both interleaved and multileaved comparison methods require item-level clicks to infer the preference for ranked lists over others, \ac{DBGD}-type algorithms cannot be extended to learn from list-level feedback. 

In contrast to \ac{DBGD}-type algorithms, we propose two \ac{OLTR} algorithms to directly learn from list-level feedback.

\paragraph{Image retrieval based on implicit feedback.}
\label{sec:imageretrieval}
There is a growing number of studies on \ac{LTR} for image retrieval that exploit user behavior. 
\citet{jain-learning-2011} use click data as a pseudo-relevance signal to train a re-ranking model and also use PCA and Gaussian Process regression to address the sparsity problem of click data in image search.
\citet{yu-learning-2015} simultaneously use visual features and click features to learn a ranking model.
\citet{ohare-leveraging-2016} extract user behavior features such as hover-through rate and demonstrate that combining these features with content features can yield significant improvements on relevance estimation compared to purely content-based features.
However, to train an \ac{LTR} framework using these features, a manually annotated dataset is needed.

In contrast to the work listed above, we propose an \ac{OLTR} method for image search that is based on list-level feedback.  To the best of our knowledge, we are the first to do so.



\tikzstyle{startstop} = [rectangle, rounded corners, text width=2cm, minimum height=1cm, text centered, draw=black, fill=blue!10, very thick]
\tikzstyle{head1} = [rectangle, rounded corners, minimum width=1cm, minimum height=1cm, text centered, draw=black, fill=green!10, very thick]
\tikzstyle{process0} = [rectangle, rounded corners, minimum width=10cm, minimum height=2cm, text centered, draw=black, fill=red!10, very thick]
\tikzstyle{process00} = [rectangle, rounded corners, minimum width=2cm, minimum height=1cm, text centered, draw=black, fill=red!40, very thick]
\tikzstyle{process1} = [rectangle, rounded corners, minimum width=10cm, minimum height=2cm, text centered, draw=black, fill=blue!10, very thick]
\tikzstyle{process10} = [rectangle, rounded corners, minimum width=2cm, minimum height=1cm, text centered, draw=black, fill=blue!40, very thick]
\tikzstyle{process2} = [rectangle, rounded corners, minimum width=10cm, minimum height=2cm, text centered, draw=black, fill=green!10, very thick]
\tikzstyle{process20} = [rectangle, rounded corners, minimum width=2.2cm, minimum height=0.5cm, text centered, draw=black, fill=green!40, very thick]
\tikzstyle{process3} = [rectangle, rounded corners, minimum width=10cm, minimum height=2cm, text centered, draw=black, fill=yellow!10, very thick]
\tikzstyle{process30} = [rectangle, rounded corners, minimum width=2cm, minimum height=1cm, text centered, draw=black, fill=yellow!40, very thick]
\tikzstyle{arrow} = [very thick,->,>=stealth]
\tikzstyle{line} = [very thick,-,>=stealth]

.

\section{Method}
\label{section:method}
In this section, we first present the notation and our general \ac{OLTR} framework (Section~\ref{sec:model}).
Then, we propose two algorithms for \ac{OLTR} with list-level feedback.
The first, \OurMethodPG, is based on policy gradients~\cite{rlsurvery} (Section~\ref{sec:pglearn}).
The second, \OurMethod, is based on linear regression  (Section~\ref{seec:reglearn}). 

\begin{algorithm}[!h]
	\begin{algorithmic}[1]
		\Input SERP size $k$ and exploration rate $\epsilon$. 
		\For{$t \leftarrow 1,2,\ldots$}	\label{alg:begin_for}
		\State $q_t \leftarrow receive\_query(t)$\label{alg:query} \Comment Receive a query from a user.
		\State $s_i \leftarrow f(\mathbf{x}_i; q_t)$ for $\forall \mathbf{x}_i \in \mathbf{X}$ \label{alg:score}\Comment Score image candidates by the ranking function $f(\mathbf{x}; q_t)$.
		\State $l_t \leftarrow generate\_results(\{s_i\}_{i=1}^n, k, \epsilon)$ \label{alg:list} \Comment Generate  ranked list $l_t$ with $\epsilon$-greedy exploration.
		\State Show $l_t$ to the user and receive the list-level feedback $r_t$. \label{alg:feedback}
		\State Update $f(\mathbf{x}; q_t)$ \label{alg:update} \Comment We update the ranking function by the proposed \OurMethodPG{} and \OurMethod{}, respectively. 
		\EndFor 
	\end{algorithmic}
	\caption{\ac{OLTR} framework}
	\label{alg:framework}
\end{algorithm}

\subsection{Notation and framework}
\label{sec:model}

In our image filtering task,  we have a set of standing queries $\mathbf{Q} = \{q_i\}_{i=1}^m$ and  a set of images $\mathbf{X} = \{\mathbf{x}_i\}_{i=1}^n$. The goal is to rank image candidates with respect to a given standing query.
We use $f(\mathbf{x}; q)$ to denote the score assigned to an image $\mathbf{x}$ by a ranking model given a query $q$,
$\theta$ to denote the parameters of the ranking function $f(\mathbf{x}; q)$,
$l$ and $r(l)$ to denote a ranked list and its list-level feedback, respectively. 

In \ac{OLTR}, a little exploration helps to increase the performance of an online algorithm~\cite{balancing}. 
But too much exploration may hurt the user experience. 
In this paper, the $\epsilon$-greedy policy is chosen to  balance exploration and exploitation.
With the $\epsilon$-greedy policy, an algorithm ranks image candidates randomly  with probability $\epsilon$ (exploration), while with probability $1-\epsilon$ the algorithm ranks image candidates based on the scores produced by the underlying \ac{DNN} (exploitation). 

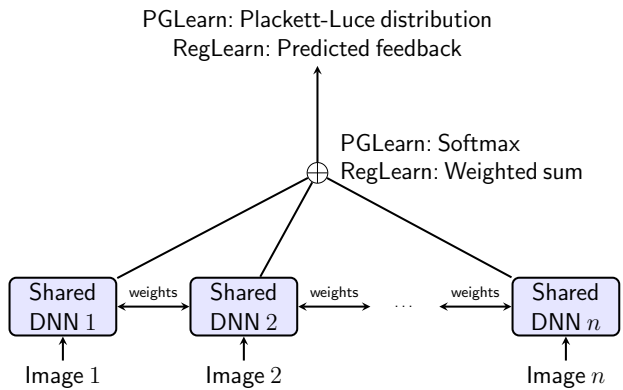
\begin{figure}[h]
 	\centering
\tikzset{XOR/.style={draw,circle,append after command={
        [shorten >=\pgflinewidth, shorten <=\pgflinewidth,]
        (\tikzlastnode.north) edge (\tikzlastnode.south)
        (\tikzlastnode.east) edge (\tikzlastnode.west)
        }
    }
}
\tikzset{line/.style={draw, -latex',shorten <=1bp,shorten >=1bp}}

\hspace*{-2mm}
\scalebox{0.64}{%
\begin{tikzpicture}[transform shape, node distance=2cm, font=\sffamily]
\node (resnet1) [startstop] {\parbox{1.5cm}{\centering\Large Shared\\ DNN $1$}};  
\node (resnet2) [startstop, right=1.5cm of resnet1] {\parbox{1.5cm}{\centering\Large  Shared\\ DNN $2$}};  
\node(dots)[right=1.5cm of resnet2, text width=1.2cm, text centered] {\ldots};
\node (resnetn) [startstop, right=1.5cm of dots] {\parbox{1.5cm}{\centering\Large Shared\\ DNN $n$}};  
\node(image1)[text width=3cm, text centered, below=0.5cm of resnet1] {\Large Image $1$};
\node(image2)[text width=3cm, text centered, below=0.5cm of resnet2] {\Large Image $2$};
\node(imagen)[text width=3cm, text centered, below=0.5cm of resnetn] {\Large Image $n$};
\node (XOR-aa)[XOR, scale=1.4, above right=2cm and 4cm of resnet1] {};
\node(sum)[text width=6cm, above right=-0.5cm and 0.2cm of XOR-aa] {\Large \OurMethodPG: Softmax \\[0.9ex] \OurMethod: Weighted sum};
\node(predict)[text width=8cm, text centered, above=2cm of XOR-aa] {\Large \OurMethodPG: Plackett-Luce distribution \\[0.9ex] \OurMethod: Predicted feedback};
\draw [->,very thick,>=stealth] (image1) -- (resnet1);
\draw [->,very thick,>=stealth] (image2) -- (resnet2);
\draw [->,very thick,>=stealth] (imagen) -- (resnetn);
\draw [<->,very thick,>=stealth] (resnet1) -- node [text width=2.5cm, midway, above, align=center ] {weights} (resnet2);
\draw [<->,very thick,>=stealth] (resnet2) -- node [text width=2.5cm, midway, above, align=center ] {weights} (dots);
\draw [<->,very thick,>=stealth] (dots) -- node [text width=2.5cm, midway, above, align=center ] {weights} (resnetn);
\draw [-,very thick] (resnet1) -- (XOR-aa);
\draw [-,very thick] (resnet2) -- (XOR-aa);
\draw [-,very thick] (resnetn) -- (XOR-aa);
\draw [->,very thick,>=stealth] (XOR-aa) -- (predict);
\end{tikzpicture}
}

\smallskip
 \caption{Structure of the proposed algorithms for \ac{OLTR} with list-level feedback.
 First, each input image receives a relevance score from the underlying ranking mode.
 Then, these scores are transformed either using softmax in \OurMethodPG\ or weighted sum in \OurMethod.
 Finally, \OurMethodPG\ outputs a Plackett-Luce distribution, while \OurMethod\ directly predicts list-level feedback.}
 \label{fig:method}
 \end{figure}

Figure~\ref{fig:method} provides a high-level overview of \OurMethodPG{} and \OurMethod{}. 
The inputs are ranked image lists. 
In this paper, both algorithms use a \acf{DNN} as the underlying ranking model to score images for a given query.
Each \ac{DNN} has $m$ outputs which is the same size as the number of standing queries. 
And a query $q$ is encoded by a one-hot vector with size $m$. 
Hence, each output of \ac{DNN} is the score of the image given the corresponding query. 
The general online learning to rank framework is provided as  pseudo-code of in Algorithm~\ref{alg:framework}.
In the rest of this section, we explain \OurMethodPG{} and \OurMethod{} in more detail, respectively. 


\subsection{\OurMethodPG{}}
\label{sec:pglearn}

The first proposed \ac{OLTR} algorithm is \OurMethodPG{}, which is based on policy gradients in reinforcement learning~\cite{rlsurvery}. 
It aims to estimate the probability of the ranked list given a query. 
Shown in Figure~\ref{fig:method}, a shared weight \ac{DNN} is employed to predict a relevance score for every image given a query. 
Then, the scores are transformed into probabilistic outputs via a softmax layer. 
\OurMethodPG{} uses these probabilities to estimate the probability distribution over actions, i.e., ranked lists of images in our case.
To get the probability of a ranked list, we follow the Plackett-Luce model~\cite{plackett2} and compute this probability as follows:
\begin{equation}
PL(l\mid q, X) = \prod_{i=1}^k \frac{\exp(f(\mathbf{x}_i; q))}{\sum_{m=i}^k \exp(f(\mathbf{x}_m; q))}.
\label{eq:plackett_luce}
\end{equation} 
%
\OurMethodPG{} identifies the best ranked list by choosing the one with the highest probability. 
However, the action space is very large, namely $O(k!)$, where $k$ is the size of a list. 
Again, we follow the Plackett-Luce model and sample without replacement from a probability distribution over the set of images, similar to~\citep{rolf-listnet}. 
Importantly, finding the best ranked list becomes infeasible as the list size increases.

In the training phase, \OurMethodPG{} learns the underlying ranking model, i.e., a \ac{DNN}, by maximizing the expected reward over the whole action space: 
\begin{equation}
\mathcal{L}(\theta) = \mathbb{E}_{q, l}[r(l, q)]. 
\label{eq:objective}
\end{equation}
 \OurMethodPG{} computes the gradients as follows: 
\begin{eqnarray}
\nabla_\theta \mathcal{L}(\theta) &=& \sum_q P(q) \sum_l \nabla_\theta \big(PL_\theta(l|q) \cdot r(l,q)\big).
 \nonumber \\
 &=& \sum_q P(q) \sum_l \frac{PL_\theta(l|q)}{PL_\theta(l|q)} \cdot \nabla_\theta PL_\theta(l|q) \cdot r(l,q) \nonumber\\
&=& \mathbb{E}_{q, l}[\nabla_\theta \log PL_\theta(l|q) \cdot r(l,q)].
\label{eq:naive_grad}
\end{eqnarray} 
Then, we approximate the derivative by sampling, for example using a single Monte Carlo sample, which is a standard procedure in policy gradients~\cite{rlsurvery}. 
Finally, we use back propagation to train the underlying \ac{DNN}.

 \subsection{\OurMethod}
 \label{seec:reglearn} 
In contract to \OurMethodPG{}, which estimates the probability of a ranked list, the second proposed algorithm, \OurMethod{}, aims at directly estimating the quality, i.e., the reward, of a ranked list given a query. 
Specifically, \OurMethod{} directly predicts the list-level feedback $r(l)$ of a ranked list $l$ for a given query $q$. 

As shown in Figure~\ref{fig:method}, given a ranked list with $n$ images,
\OurMethod{} employs $n$ \ac{DNN}s, each of which shares the weights, to score all the images. 
Then, \OurMethod{} sums up the scores output by \ac{DNN}s with some discounted weights to approximate the feedback $\hat{r}(l)$. 
More precisely, \OurMethod{}  approximates the feedback as follows: $\hat{r}(l) = \sum_{i=1}^{n}  {w}_i f(\mathbf{x}_i, q)$, where $\mathbf{w} = \{w_1, \ldots, w_n\}$ are the discounted weights of positions. 
These $\mathbf{w}$ are regarded as the last layer (the discounted summation lay) in Figure~\ref{fig:method}, and are learned during training.  
Finally, the $L2$-loss is used to optimize the underlying \ac{DNN} and the above-mentioned weights: 
\begin{equation}
\label{eq:dcgloss}
	\mathcal{L}(r(l), \hat{r}(l)) = \frac{1}{2}(r(l)-\hat{r}(l))^2. 
\end{equation} 
Since the discounted weights $\mathbf{w}$ just mentioned are the last layer of the whole network structure and  the L2-loss is differentiable, the error can be  back propagated to the underlying \ac{DNN}. 


\section{Experimental setup}
\label{section:experimental-setup}

Our experiments are designed to answer three research questions: 
\begin{inparaenum}
	\item[(\textbf{RQ1})] 
	Can  \OurMethod{} and \OurMethodPG{} learn from the ideal feedback, i.e. nDCG@$k$? 
	\item[(\textbf{RQ2})]  How would different levels of noise in the feedback signal affect the performance of  \OurMethod{}? 
	\item[(\textbf{RQ3})]  Can \OurMethod{} learn the discounted weights $\mathbf{w}$ while learning the ranking function?  
\end{inparaenum}

\paragraph{Dataset.}
We conduct experiments on the MSCOCO image dataset~\cite{mscoco}.
In the MSCOCO dataset, each image contains at least one object, where an object can be seen as a category the image belongs to or a query the image is relevant to, and so an image is relevant to one or more queries.
More precisely, MSCOCO contains $2.5$ million labeled objects in $328$k images chosen from a set of $80$ objects.
In our image filtering setup, this translates into $80$ standing queries and $328$k images to rank.
We train our models on the training set of MSCOCO.
Since the test set of MSCOCO does not have labels, we test our methods on the validation set of MSCOCO. 

\paragraph{Online learning simulation.}
Evaluating the ability of an online algorithm requires a sequence of user requests (queries) and user feedback. 
The ideal setup is to run algorithms on the real online systems and interact with real users. 
However, online experiments are expensive. 
In contrast, simulation experiments, which are cheaper, are widely used to evaluate the ability of online algorithms~\cite{kveton2018bubblerank,hofmann-fidelity-2013,hofmann-reusing-2013,balancing,chang2018mergedts}. 
In this paper, we use the following procedure to simulate \ac{OLTR}: 
\begin{enumerate}[nosep]
    \item Given a randomly chosen standing query, a set of candidate images is selected randomly, such that at least one of the selected images is relevant to the query. 
    This is a general setup in \ac{OLTR}~\cite{batchrank,kveton2018bubblerank}. 
    \item We use ResNet~\cite{resnet} to produce a relevance score for each selected image. Before passing an image to ResNet, we apply VGG-Net preprocessing to rescale the image and obtain random crops so as to avoid overfitting~\citep{simonyan-very-2014}.
    \item The list-level feedback of the ranked list is then simulated. 
    \item The user feedback is used by \OurMethodPG\ and \OurMethod\ to update the ranking model, namely ResNet.
\end{enumerate}

\noindent%
Since we focus on the image filtering scenario with a finite set of queries, the same query appears multiple times in the training and test set.
Every image appears in only one of the two sets (training or test). 

In order to avoid updating  ResNet frequently and to take advantage of parallelized computations, the above procedure is performed in batches. 
Each batch contains $100$ queries, which are processed in parallel.
We update ResNet after we collect feedback for all $100$ queries in a batch. 
In terms of hyperparameters, we use a learning rate of $10^{-4}$ together with the Adam optimizer~\cite{adam}. 
We also use batch normalization~\cite{batchnormalization} with a decay of $0.997$ and epsilon of $10^{-5}$ which are the default setups in ResNet~\cite{resnet}.

\paragraph{List-level feedback.}
We use two types of list-level feedback: nDCG@$k$ and CTR@$k$, where $k$ is the number of positions. 
nDCG@$k$ is a widely used metric in ranking tasks~\cite{balancing,mdgd,hofmann-reusing-2013}; it measures the quality of a ranked list. 
It is an ideal and deterministic feedback for \ac{OLTR} but it is not clear how to transfer real world feedback to nDCG scores. 
We conduct experiments with nDCG@$k$ feedback to determine whether \OurMethod{} and \OurMethodPG{} are able to learn from list-level feedback. 
Obviously, if they could not learn from nDCG@$k$ feedback, they certainly would not learn from noisy feedback, which widely exists in online interactive search systems~\cite{dbgd,lambdaclick}. 
For a more realistic setup, we choose CTR@$k$ as another type of list-level feedback, which is easily obtained from users and contains more noise. 
In experiments, we choose $k=2$ and $5$, which are important in commercial search systems~\cite{lambdaclick,Williams:top5}. 

\begin{table*}[t]
	\centering
	\caption{Overview of the click configurations. 
	}
	\label{tb:transfer}
	\begin{tabular}{p{2.3cm}ll}
		\toprule	
		configuration & $p(a|R)$ & $p(a|IR)$ \\
		\midrule
		perfect & $1.0$ & $0.0$ \\
		locating & $0.95$ & $0.05$ \\
		entertaining & $0.9$ & $0.4$ \\
		\bottomrule
	\end{tabular}
\end{table*}

\paragraph{Click simulation.}
For the click feedback based experiments, we use the Position Based Model (PBM)~\cite{click-2015} to simulate clicks. 
This configuration is different from the Dependent Click Model (DCM)~\cite{click-2015} based configurations, i.e., perfect, navigational and informational configurations~\cite{balancing}, which are widely used in the online text retrieval simulation. 
The reason that we choose PBM instead of DCM in the paper is twofold: 
\begin{inparaenum}
	\item Recent user studies show that users do not always check the highest ranked images~\cite{wang2018large,xie-2018-image,xie-investigating-2017},  so DCM does not exactly fit for the image  retrieval task. 
	\item The image retrieval scenario is more complex than the text retrieval scenario~\cite{xie-investigating-2017,xie-why-2018,wang2018large}. Particularly, there is no certain order in positions with which users browse a search engine result page. 
\end{inparaenum}

The PBM consists of two sets of parameters: examination and attraction probabilities~\cite{click-2015}. 
For the examination probabilities, we learn the PBM of the \emph{Yandex} click logs\footnote{\url{ https://academy.yandex.ru/events/data\_analysis/relpred2011}} using PyClick\footnote{\url{https://github.com/markovi/PyClick}} and obtain the examination probabilities of the top $5$ positions as follows: $0.999, 0.959, 0.761, 0.592$ and  $0.457$.  

For the attraction probabilities, we follow the configurations used by \citet{balancing} and design three configurations, i.e., \emph{perfect}, \emph{locating} and \emph{entertaining}, to  transfer the relevance labels in the MSCOCO dataset into attraction probabilities.
Table~\ref{tb:transfer} provides an overview of the configurations. 
The feedback in the perfect configuration is deterministic, where users always and only click relevant images.
This configuration aims at upper bounding the performance with click feedback. 
The other configurations are designed to mimic two types of user behavior in image search, i.e., ``locate" and ``entertain," as proposed by~\citet{xie-why-2018}.
In the locating configuration, a user tries to find certain images that match some requirements, so he or she has  a high probability to click relevant images and a low probability to click irrelevant images. 
In the entertaining configurations, instead of finding certain images, users want to kill time by browsing the search results, so their click behaviors tend to be noisy. 
All in all, the noise level in three configurations follows this order: perfect" $<$ ``locating" $<$ ``entertaining."

\paragraph{Skyline.}
Since all previous \ac{OLTR} algorithms require item-level feedback, we do not compare \OurMethodPG{} and \OurMethod{} with any baseline.
However, to calibrate their performance, we design a \emph{skyline} for the comparison. 
The skyline we choose is based on \OurMethod{}, but the last layer is the predefined discounted weights and not updated during training. 
More precisely, for the different types of list-level feedback,  either the discounted weights in nDCG or the examination probability of a PBM are input to \OurMethod{} as prior knowledge.
Since \OurMethod{} is informed about the importance of each position in this way,  it should learn a high quality ranking model. 
We call the skyline {\skyline{}}. 

\paragraph{Exploration.}
To  explore the ranking space, all the algorithms are combined with 
two types of $\epsilon$-greedy policy with $\epsilon = 0.1$ and $\epsilon = 1$. 
The setting $\epsilon=1$ means that the algorithm always explores the ranking space, which may lead to the best offline performance. 
We also choose $\epsilon=0.1$ because little exploration is necessary for online algorithms and $\epsilon=0.1$ is a good choice for \ac{OLTR} algorithms~\cite{balancing}. 
So, in our experiments, we have $6$ combinations of learning algorithms and exploration strategies: \OurMethodPG{}, \OurMethod{}, and \skyline, each with $\epsilon =0.1$ and $\epsilon = 1$.

\paragraph{Evaluation measures.}
We have two types of evaluation: online and offline.
In online evaluation, we care about the user experience during the whole online training phase. 
Both the past and  current performances of the algorithm should be considered when choosing the online metric. 
We use the average cumulative nDCG@$k$ of ranked lists as the online metric, which is computed as follows: 
$nDCG@k (T) = \frac{1}{T} \sum_{t=1}^{T}nDCG@k(l_t)$, where $T$ is the number of steps and $nDCG@k(l_t)$ is the nDCG@$k$ of the ranked list $l_t$. 

In offline evaluation, we care about the final quality of \ac{OLTR} algorithms, which is measured by nDCG@$k$ on a left out test set.
To measure the offline performance, we randomly choose $150$ batches each containing $100$ queries from the test set. 
Then, we compute the average nDCG@$k$ over these $15,000$ queries. 
To determine whether differences between the best performance and the others are significant, 
we use  a two-tailed Student's t-test with $p < 0.05$.


\section{Results}
\label{section:results}
In this section, we report and analyze the results of the experiments described in Section~\ref{section:experimental-setup}.

\begin{figure*}[t]
	\centering
	\includegraphics[width = \textwidth]{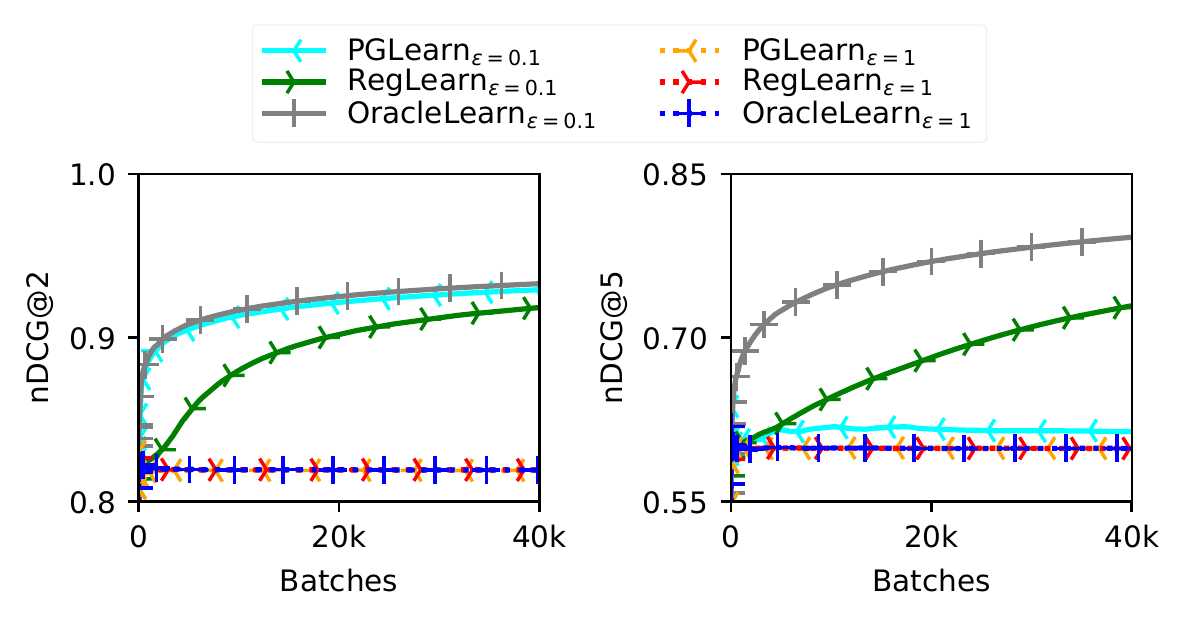}	
	\caption{Online performance (average nDCG$@k$) of different combinations of learning methods (\OurMethodPG{}, \OurMethod{} and \skyline{}) and exploration rate ($\epsilon =0.1$ and $\epsilon =1$). Higher is better. Best viewed in color.}  
	\label{fig:performance}	
\end{figure*}
 
\paragraph{Online performance with nDCG feedback.}
\label{sec:results_loss}
Figure~\ref{fig:performance} presents the average rewards (average nDCG@$k$) acquired by \OurMethodPG{}, \OurMethod{} and \skyline\ combining with $\epsilon$-greedy with $\epsilon=0.1$ and $\epsilon=1$ during the online training. 
We run experiments with $40$k batches, which contain $4$M queries in total. 

We see that all algorithms with $\epsilon=1$ (pure exploration) have the worst performance since the output  is just a randomly ranked list. 
\skyline$_{\epsilon=0.1}$ outperforms others in both the $k=2$ and $k=5$ cases. 
For $k=2$, \OurMethodPG$_{\epsilon=0.1}$ performs close to \skyline$_{\epsilon=0.1}$ and outperforms \OurMethod$_{\epsilon=0.1}$. 
But when $k$ becomes larger, i.e., for $k=5$, \OurMethodPG$_{\epsilon=0.1}$ is hardly better than generating random output and loses to \OurMethod$_{\epsilon=0.1}$ with a large gap. 
The huge difference in the behavior of \OurMethodPG$_{\epsilon=0.1}$ in the two cases is caused by the fact that policy gradient tends to converge to suboptimal policies without a good estimate of the reward~\cite{rlsurvery}.  
We approximate the derivate (Eq.~\ref{eq:naive_grad}) by Monte Carlo sampling. 
When the action space is small, e.g., $O(2!)$ for $k=2$, Monte Carlo sampling can easily sample all actions and then return a good estimation of the rewards, so we get a good approximation of the derivate. 
But when the action space is large, e.g., $O(5!)$ for $k=5$, Monte Carlo sampling  can hardly sample the whole action space and then the estimation of rewards is biased and the approximation of the derivate is inaccurate. 
Given the fact that the SERP size can be larger than $5$, where \OurMethodPG$_{\epsilon=0.1}$ may perform even worse,   \OurMethod$_{\epsilon=0.1}$ is a better choice for the online image filtering task than \OurMethodPG$_{\epsilon=0.1}$ and we omit $\OurMethodPG{}$ from further considerations.

\paragraph{Offline performance.}

\begin{table*}[t]
	\centering
	\caption{Offline  nDCG@$k$ scores obtained on the test set of MSCOCO. The best results are marked in boldface. 
		Statistically significant losses against the best result per row are indicated by $^*$. 
		We report the standard deviation in the subscripts. 
	}
	\label{tb:results_new}
		\begin{tabular}{lllllll}
			\toprule	
			& \multicolumn{2}{c}{\OurMethodPG{} }&
			\multicolumn{2}{c}{\OurMethod{} }& \multicolumn{2}{c}{\skyline{} (\emph{skyline})} \\			
			\cmidrule(lr){2-3}\cmidrule(lr){4-5} \cmidrule(lr){6-7}  
			&$\epsilon=1$  &$\epsilon=0.1$
			&$\epsilon=1$  &$\epsilon=0.1$
			&$\epsilon=1$  &$\epsilon=0.1$  \\
			
			\midrule
			
			$k=2$
			&$0.926_{0.02}^*$&$0.946_{0.02}^*$&$0.846_{0.13}^*$&$\mathbf{0.955}_{0.02}$&$\mathbf{0.955}_{0.02}$&$0.952_{0.02}$ \\
			$k=5$
			&$0.607_{0.05}^*$&$0.624_{0.05}^*$&$0.564_{0.18}^*$&$0.825_{0.05}^*$&$0.814_{0.05}^*$&$\mathbf{0.840}_{0.05}$ \\
			\bottomrule
	\end{tabular}
%
%
\end{table*}

\hyphenation{Oracle-Learn}

We report the offline performance of the algorithms trained with  nDCG@$k$ feedback in Table~\ref{tb:results_new}. 
The values reflect the quality of  user experiences after the training. 
For $k=2$,  \skyline$_{\epsilon=1}$ and \OurMethod$_{\epsilon=0.1}$ have the best offline performance. 
Different from the online results, where  \OurMethodPG$_{\epsilon=0.1}$ outperforms \OurMethod$_{\epsilon=0.1}$, \OurMethod$_{\epsilon=0.1}$ outperforms PG\-Learn$_{\epsilon=0.1}$ significantly with the offline metric.  
We hypothesize  that even for $k=2$ the action space is too large for the Monte Carlo sampling, used by \OurMethodPG{}, to fully estimate the action reward and the good online performance may be the effect of overfitting the training dataset. 
For a longer list, e.g., with $k=5$, \skyline$_{\epsilon=0.1}$ outperforms others.  \OurMethod$_{\epsilon=0.1}$ does not keep up with Oracle\-Learn$_{\epsilon=0.1}$.  
That is, \OurMethod$_{\epsilon=0.1}$ is about $0.015$ lower  than \skyline$_{\epsilon=0.1}$   in term of nDCG@$5$. 
Again, \OurMethodPG\ performs poorly. 

The offline evaluation indicates that, with the proper exploration policy, \OurMethod{} can learn to combine the individual scores to predict list-level feedback, and then train a good ranking model, especially with short lists, e.g. $k=2$.  
However, pure exploration, i.e., $\epsilon=1$, is harmful to \OurMethod{}.  
This is because \OurMethod\ cannot get any position information through pure exploration, which randomly shuffles the results during the training phase. 
On the other hand, \OurMethodPG{} performs poorly when $k=5$, which is consistent with the online results. 
Together with the online performance, the answer to \textbf{RQ1} is that \OurMethod{} can learn from list-level feedback and  \OurMethodPG{} can learn from the list-level feedback with small list sizes. 
Because of the poor performance of \OurMethodPG{} when $k>2$, we leave it out of our later experiment with click feedback.  

\paragraph{Online performance with click feedback.}
To answer \textbf{RQ2}, we conduct  experiments on \OurMethod$_{\epsilon=0.1}$  and \skyline$_{\epsilon=0.1}$  with click feedback. 
Here, \skyline\ has the prior knowledge of the examination probabilities. 
We omit the results of $k=2$, because the difference between the examination probabilities of the first two positions is fairly small, only $0.04$.  
We choose $\epsilon$-greedy with $\epsilon=0.1$, since pure exploration ($\epsilon=1$) hurts the performance of \OurMethod{}.

\begin{figure*}[t]
	\centering
	\includegraphics[width = \textwidth]{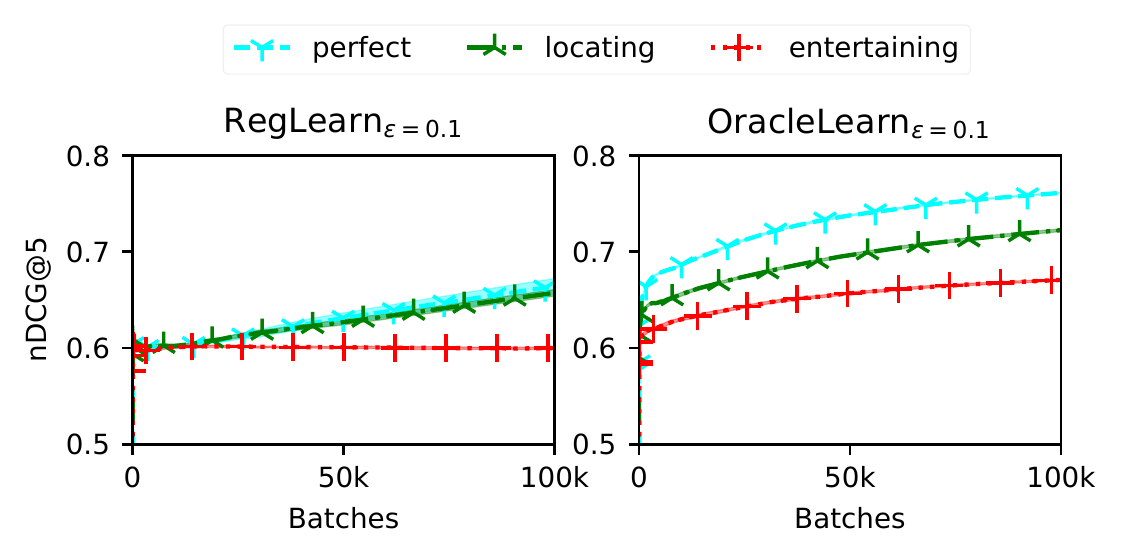}
	\caption{Online performances (average nDCG$@5$) of \OurMethod$_{\epsilon=0.1}$ and \skyline$_{\epsilon=0.1}$ with three different click configurations.  
	Higher is better.
	Best viewed in color.}  
	\label{fig:click}	
\end{figure*}

Shown in Fig.~\ref{fig:click}, with the perfect and locating configurations, the performances (average nDCG@$5$) of \OurMethod$_{\epsilon=0.1}$ go from $0.59 \pm 0.01$ and $0.61 \pm 0.02$ to $0.67 \pm 0.01$ and $0.66 \pm 0.00$, respectively.
However, with the entertaining configuration, \OurMethod$_{\epsilon=0.1}$ fails to learn a ranking function. 
Note that the entertaining configuration is the most noisy one. 
This result indicates that \OurMethod$_{\epsilon=0.1}$  can learn from  noisy feedback,  but the performance of  \OurMethod$_{\epsilon=0.1}$  drops down as the level of noise increases. 
If the feedback is too noisy, as witnessed in, e.g., the entertaining configuration,  \OurMethod$_{\epsilon=0.1}$ may fail to learn a ranking model. 

When it comes to the \skyline$_{\epsilon=0.1}$, we see that it learns a ranking function with all three configurations, since the three lines in the right plot of Figure~\ref{fig:click} climb up alone the number of batches. 
This result  indicates that the performances of  \OurMethod{} can be boosted by integrating the  prior knowledge of user behavior. 

\begin{figure*}[t]
	\centering
	\includegraphics[width = .49\textwidth]{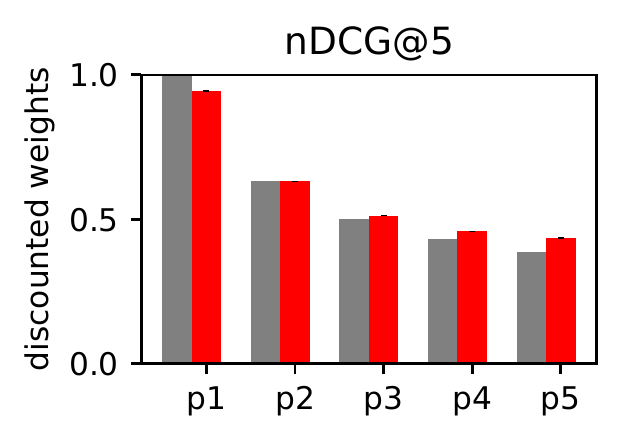}
	\includegraphics[width = .49\textwidth]{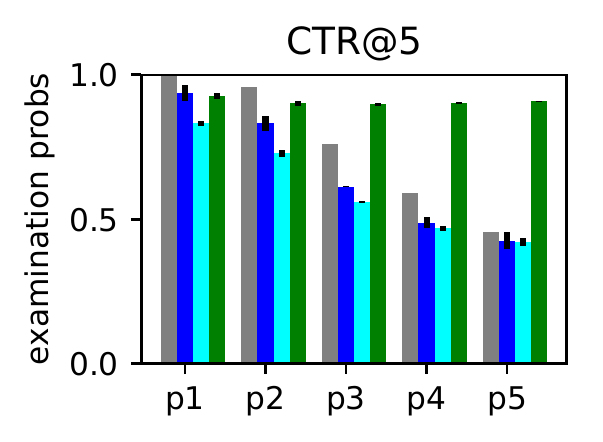}
	\caption{(Left): the ground truth (gray) and the discounted weight of each position learned by \OurMethod$_{\epsilon=0.1}$ (red) with nDCG$@5$ feedback. 
		(Right):  the ground truth (gray) and the examination probability  of each position learned by \OurMethod$_{\epsilon=0.1}$ with perfect (Blue), locating (cyan) and entertaining (green) click configurations. 
		Best viewed in color.}
	\label{fig:weights}	
\end{figure*}

\paragraph{Learning discounted weights.}
To answer \textbf{RQ3}, we analyze the  discounted  weights, $\mathbf{w}$, learned by \OurMethod$_{\epsilon=0.1}$, i.e., the discounted weights in nDCG and the examination probabilities in a PBM. 
Figure~\ref{fig:weights} shows the results with different setups, where 
 the ``ground truth" is the true $\mathbf{w}$ in nDCG and click setup, respectively. 

For the nDCG feedback setup, the learned $\mathbf{w}$ by \OurMethod$_{\epsilon=0.1}$  is close to the ground truth, and  the  Euclidean distance\footnote{We consider the learned weights as a $5$-dimensional vector.} to the ground truth is  $0.079$. 
When it comes to the click feedback setup, with the less noisy configurations, i.e., the perfect and locating configurations,  \OurMethod$_{\epsilon=0.1}$ learns the correct order of the importance of positions. 
However, because of  the noise in feedback, the Euclidean distances from  learned $\mathbf{w}$ to the ground truth  are $0.231$ and $0.372$, respectively. 
Moreover, \OurMethod$_{\epsilon=0.1}$ fails to learn the examination probabilities with the entertaining configuration, the most noisy configuration.

In summary, the answer to \textbf{RQ3} is that  \OurMethod$_{\epsilon=0.1}$ can learn the proper discounted weight $\mathbf{w}$ from the ideal or little noisy feedback, but cannot learn a proper $\mathbf{w}$ from the most noisy feedback, i.e.,  entertaining click feedback. 
%


\section{Conclusion}
\label{section:conclusion}

This paper has shown two novel ways to use list-level feedback for \ac{OLTR}: \OurMethodPG{} and \OurMethod{}.  
They can both back propagate the loss to the underlying ranking model. 
The main findings are that \OurMethod{} with $\epsilon$-greedy, with $\epsilon=0.1$, is a good choice for the \ac{OLTR} task in the image filtering setup. 

One of the interesting future directions is to adopt different types of exploration strategy to \OurMethodPG{} and \OurMethod{}, since the experiments have demonstrated that proper exploration helps to boost offline performance, while, with pure exploration, $\epsilon=1$, \OurMethod{} fails to train the underlying ranking model. 
We hypothesize that an alternative exploration policy may help to increase the performance of \OurMethod{}.

\subsection*{Code and data}
To facilitate reproducibility of the results in this paper, we are sharing the code and data used to run the experiments in this paper at \url{https://github.com/chang-li/dcgnet}.

\bibliographystyle{abbrvnatnourl}
\bibliography{references} 

\begin{thebibliography}{38}
\providecommand{\natexlab}[1]{#1}
\providecommand{\url}[1]{\texttt{#1}}
\expandafter\ifx\csname urlstyle\endcsname\relax
  \providecommand{\doi}[1]{doi: #1}\else
  \providecommand{\doi}{doi: \begingroup \urlstyle{rm}\Url}\fi

\bibitem[Amig{\'o} et~al.(2014)Amig{\'o}, Carrillo-de Albornoz, Chugur, Corujo,
  Gonzalo, Meij, de~Rijke, and Spina]{amigo-overview-2014}
E.~Amig{\'o}, J.~Carrillo-de Albornoz, I.~Chugur, A.~Corujo, J.~Gonzalo,
  E.~Meij, M.~de~Rijke, and D.~Spina.
\newblock Overview of {RepLab} 2014: Author profiling and reputation dimensions
  for online reputation management.
\newblock In \emph{CLEF}, pages 307--322. Springer, 2014.

\bibitem[Belkin and Croft(1992)]{filtering1}
N.~J. Belkin and W.~B. Croft.
\newblock Information filtering and information retrieval: Two sides of the
  same coin?
\newblock \emph{Communications of ACM}, 35\penalty0 (12):\penalty0 29--38,
  1992.

\bibitem[Boom et~al.(2014)Boom, He, et~al.]{boom-research-2014}
B.~J. Boom, J.~He, et~al.
\newblock A research tool for long-term and continuous analysis of fish
  assemblage in coral-reefs using underwater camera footage.
\newblock \emph{Ecological Informatics}, 23:\penalty0 83--97, 2014.

\bibitem[Chuklin et~al.(2015)Chuklin, Markov, and de~Rijke]{click-2015}
A.~Chuklin, I.~Markov, and M.~de~Rijke.
\newblock \emph{Click Models for Web Search}.
\newblock Synthesis Lectures on Information Concepts, Retrieval, and Services.
  Morgan \& Claypool Publishers, August 2015.

\bibitem[DeClaris et~al.(1994)DeClaris, Harman, Faloutsos, Dumais, and
  Oard]{filtering3}
N.~DeClaris, D.~K. Harman, C.~Faloutsos, S.~Dumais, and D.~Oard.
\newblock Information filtering and retrieval: overview, issues and directions.
\newblock In \emph{EMBC}, volume~1, pages A42--A49, 1994.

\bibitem[He et~al.(2016)He, Zhang, Ren, and Sun]{resnet}
K.~He, X.~Zhang, S.~Ren, and J.~Sun.
\newblock Deep residual learning for image recognition.
\newblock In \emph{CVPR}, pages 770--778, 2016.

\bibitem[Henery(1981)]{plackett2}
R.~Henery.
\newblock Permutation probabilities as models for horse races.
\newblock \emph{Journal of the Royal Statistical Society. Series B
  (Methodological)}, 43\penalty0 (1):\penalty0 86--91, 1981.

\bibitem[Hofmann et~al.(2013{\natexlab{a}})Hofmann, Schuth, Whiteson, and
  de~Rijke]{hofmann-reusing-2013}
K.~Hofmann, A.~Schuth, S.~Whiteson, and M.~de~Rijke.
\newblock Reusing historical interaction data for faster online learning to
  rank for information retrieval.
\newblock In \emph{WSDM}. ACM, February 2013{\natexlab{a}}.

\bibitem[Hofmann et~al.(2013{\natexlab{b}})Hofmann, Whiteson, and
  de~Rijke]{balancing}
K.~Hofmann, S.~Whiteson, and M.~de~Rijke.
\newblock Balancing exploration and exploitation in listwise and pairwise
  online learning to rank for information retrieval.
\newblock \emph{Information Retrieval Journal}, 16\penalty0 (1):\penalty0
  63--90, February 2013{\natexlab{b}}.

\bibitem[Hofmann et~al.(2013{\natexlab{c}})Hofmann, Whiteson, and
  de~Rijke]{hofmann-fidelity-2013}
K.~Hofmann, S.~Whiteson, and M.~de~Rijke.
\newblock Fidelity, soundness, and efficiency of interleaved comparison
  methods.
\newblock \emph{ACM Transactions on Information Systems}, 31\penalty0
  (3):\penalty0 Article 18, October 2013{\natexlab{c}}.

\bibitem[Ioffe and Szegedy(2015)]{batchnormalization}
S.~Ioffe and C.~Szegedy.
\newblock Batch normalization: Accelerating deep network training by reducing
  internal covariate shift.
\newblock In \emph{ICML}, pages 448--456, 2015.

\bibitem[Jagerman et~al.(2017)Jagerman, Kiseleva, and de~Rijke]{rolf-listnet}
R.~Jagerman, J.~Kiseleva, and M.~de~Rijke.
\newblock Modeling label ambiguity for listwise neural learning to rank.
\newblock In \emph{Neu-IR 2017}, August 2017.

\bibitem[Jain and Varma(2011)]{jain-learning-2011}
V.~Jain and M.~Varma.
\newblock Learning to re-rank: query-dependent image re-ranking using click
  data.
\newblock In \emph{WWW}, pages 277--286. ACM, 2011.

\bibitem[Joachims et~al.(2005)Joachims, Granka, Pan, Hembrooke, and
  Gay]{joachims2017accurately}
T.~Joachims, L.~Granka, B.~Pan, H.~Hembrooke, and G.~Gay.
\newblock Accurately interpreting clickthrough data as implicit feedback.
\newblock In \emph{SIGIR}, pages 154--161. ACM, 2005.

\bibitem[Kaelbling et~al.(1996)Kaelbling, Littman, and Moore]{rlsurvery}
L.~P. Kaelbling, M.~L. Littman, and A.~W. Moore.
\newblock Reinforcement learning: A survey.
\newblock \emph{J. Artif. Intell. Res.(JAIR)}, 4:\penalty0 237--285, 1996.

\bibitem[Kingma and Ba(2014)]{adam}
D.~Kingma and J.~Ba.
\newblock Adam: A method for stochastic optimization.
\newblock \emph{arXiv preprint arXiv:1412.6980}, 2014.

\bibitem[Kiseleva et~al.(2016)Kiseleva, Williams, Hassan~Awadallah, Crook,
  Zitouni, and Anastasakos]{kiseleva2016predicting}
J.~Kiseleva, K.~Williams, A.~Hassan~Awadallah, A.~C. Crook, I.~Zitouni, and
  T.~Anastasakos.
\newblock Predicting user satisfaction with intelligent assistants.
\newblock In \emph{SIGIR}, pages 45--54. ACM, 2016.

\bibitem[Kveton et~al.(2015)Kveton, Szepesvari, Wen, and
  Ashkan]{kveton2015cascading}
B.~Kveton, C.~Szepesvari, Z.~Wen, and A.~Ashkan.
\newblock Cascading bandits: Learning to rank in the cascade model.
\newblock In \emph{ICML}, pages 767--776, 2015.

\bibitem[Kveton et~al.(2018)Kveton, Li, Lattimore, Markov, de~Rijke,
  Szepesvari, and Zoghi]{kveton2018bubblerank}
B.~Kveton, C.~Li, T.~Lattimore, I.~Markov, M.~de~Rijke, C.~Szepesvari, and
  M.~Zoghi.
\newblock Bubblerank: Safe online learning to rerank.
\newblock \emph{arXiv preprint arXiv:1806.05819}, 2018.

\bibitem[Li and de~Rijke(2018)]{li-incremental-2018}
C.~Li and M.~de~Rijke.
\newblock Incremental sparse bayesian ordinal regression.
\newblock \emph{Neural Networks}, 106:\penalty0 294--302, October 2018.

\bibitem[Li et~al.(2018)Li, Markov, de~Rijke, and Zoghi]{chang2018mergedts}
C.~Li, I.~Markov, M.~de~Rijke, and M.~Zoghi.
\newblock Merge double thompson sampling for large scale online ranker
  evaluation.
\newblock \emph{arXiv preprint arXiv:1812.04412}, 2018.

\bibitem[Lin et~al.(2014)Lin, Maire, Belongie, Hays, Perona, Ramanan,
  Doll{\'a}r, and Zitnick]{mscoco}
T.-Y. Lin, M.~Maire, S.~Belongie, J.~Hays, P.~Perona, D.~Ramanan,
  P.~Doll{\'a}r, and C.~L. Zitnick.
\newblock Microsoft {COCO}: Common objects in context.
\newblock In \emph{ECCV}, pages 740--755, 2014.

\bibitem[Liu(2009)]{ltr}
T.-Y. Liu.
\newblock Learning to rank for information retrieval.
\newblock \emph{Foundations and Trends in Information Retrieval}, 3\penalty0
  (3):\penalty0 225--331, 2009.

\bibitem[O'Hare et~al.(2016)O'Hare, de~Juan, Schifanella, He, Yin, and
  Chang]{ohare-leveraging-2016}
N.~O'Hare, P.~de~Juan, R.~Schifanella, Y.~He, D.~Yin, and Y.~Chang.
\newblock Leveraging user interaction signals for web image search.
\newblock In \emph{SIGIR}, pages 559--568. ACM, 2016.

\bibitem[Radlinski et~al.(2008)Radlinski, Kleinberg, and Joachims]{rankbandit}
F.~Radlinski, R.~Kleinberg, and T.~Joachims.
\newblock Learning diverse rankings with multi-armed bandits.
\newblock In \emph{ICML}, pages 784--791. ACM, 2008.

\bibitem[Schuth et~al.(2016)Schuth, Oosterhuis, Whiteson, and de~Rijke]{mdgd}
A.~Schuth, H.~Oosterhuis, S.~Whiteson, and M.~de~Rijke.
\newblock Multileave gradient descent for fast online learning to rank.
\newblock In \emph{WSDM}, pages 457--466. ACM, February 2016.

\bibitem[Simonyan and Zisserman(2014)]{simonyan-very-2014}
K.~Simonyan and A.~Zisserman.
\newblock Very deep convolutional networks for large-scale image recognition.
\newblock In \emph{arXiv preprint arXiv:1409.1556}, 2014.

\bibitem[Wang et~al.(2018{\natexlab{a}})Wang, Langley, Kim, McCord-Snook, and
  Wang]{wang2018efficient}
H.~Wang, R.~Langley, S.~Kim, E.~McCord-Snook, and H.~Wang.
\newblock Efficient exploration of gradient space for online learning to rank.
\newblock \emph{arXiv preprint arXiv:1805.07317}, 2018{\natexlab{a}}.

\bibitem[Wang et~al.(2018{\natexlab{b}})Wang, Su, He, Liu, and
  Ma]{wang2018large}
X.~Wang, N.~Su, Z.~He, Y.~Liu, and S.~Ma.
\newblock A large-scale study of mobile search examination behavior.
\newblock In \emph{SIGIR}. ACM, July 2018{\natexlab{b}}.

\bibitem[Williams et~al.(2016)Williams, Kiseleva, Crook, Zitouni, Awadallah,
  and Khabsa]{Williams:top5}
K.~Williams, J.~Kiseleva, A.~C. Crook, I.~Zitouni, A.~H. Awadallah, and
  M.~Khabsa.
\newblock Detecting good abandonment in mobile search.
\newblock In \emph{WWW '16}, pages 495--505, 2016.

\bibitem[Xie et~al.(2017)Xie, Liu, Wang, Wang, Wu, Wu, Zhang, and
  Ma]{xie-investigating-2017}
X.~Xie, Y.~Liu, X.~Wang, M.~Wang, Z.~Wu, Y.~Wu, M.~Zhang, and S.~Ma.
\newblock Investigating examination behavior of image search users.
\newblock In \emph{SIGIR}, pages 275--284. ACM, 2017.

\bibitem[Xie et~al.(2018{\natexlab{a}})Xie, Liu, de~Rijke, He, Zhang, and
  Ma]{xie-why-2018}
X.~Xie, Y.~Liu, M.~de~Rijke, J.~He, M.~Zhang, and S.~Ma.
\newblock Why people search for images using web search engines.
\newblock In \emph{WSDM}. ACM, February 2018{\natexlab{a}}.

\bibitem[Xie et~al.(2018{\natexlab{b}})Xie, Mao, de~Rijke, Zhang, Zhang, and
  Ma]{xie-2018-image}
X.~Xie, J.~Mao, M.~de~Rijke, R.~Zhang, M.~Zhang, and S.~Ma.
\newblock Constructing an interaction behavior model for web image search.
\newblock In \emph{SIGIR}, pages 425--434. ACM, 2018{\natexlab{b}}.

\bibitem[Yu et~al.(2015)Yu, Tao, Wang, and Rui]{yu-learning-2015}
J.~Yu, D.~Tao, M.~Wang, and Y.~Rui.
\newblock Learning to rank using user clicks and visual features for image
  retrieval.
\newblock \emph{IEEE Transactions on Cybernetics}, 45\penalty0 (4):\penalty0
  767--779, April 2015.

\bibitem[Yue and Joachims(2009)]{dbgd}
Y.~Yue and T.~Joachims.
\newblock Interactively optimizing information retrieval systems as a dueling
  bandits problem.
\newblock In \emph{ICML}, pages 1201--1208. ACM, 2009.

\bibitem[Yue et~al.(2010)Yue, Patel, and Roehrig]{yue2010beyond}
Y.~Yue, R.~Patel, and H.~Roehrig.
\newblock Beyond position bias: Examining result attractiveness as a source of
  presentation bias in clickthrough data.
\newblock In \emph{WWW}, pages 1011--1018. ACM, 2010.

\bibitem[Zoghi et~al.(2016)Zoghi, Tunys, Li, Jose, Chen, Chin, and
  de~Rijke]{lambdaclick}
M.~Zoghi, T.~Tunys, L.~Li, D.~Jose, J.~Chen, C.~M. Chin, and M.~de~Rijke.
\newblock Click-based hot fixes for underperforming torso queries.
\newblock In \emph{SIGIR}, pages 195--204. ACM, July 2016.

\bibitem[Zoghi et~al.(2017)Zoghi, Tunys, Ghavamzadeh, Kveton, Szepesvari, and
  Wen]{batchrank}
M.~Zoghi, T.~Tunys, M.~Ghavamzadeh, B.~Kveton, C.~Szepesvari, and Z.~Wen.
\newblock Online learning to rank in stochastic click models.
\newblock In \emph{ICML}, pages 4199--4208, 2017.

\end{thebibliography}
 
\end{document}